%% file: emnlp2023.tex
\documentclass[10pt, a4paper]{article}

\usepackage[]{lrec-coling2024} 

\usepackage{latexsym}
\usepackage{multirow}
\usepackage{makecell}

\definecolor{midnightgreen}{rgb}{0.0, 0.29, 0.33}

\usepackage{amsmath}
\usepackage{subcaption}
\usepackage{amsfonts}
\title{Fusion-in-T5: Unifying Variant Signals for Simple and Effective Document Ranking with Attention Fusion}

\name{Shi Yu$^{1}${$^\dag$}\thanks{{$^\dag$}~~Equal contribution.}, Chenghao Fan$^{2}${$^\dag$}, Chenyan Xiong$^{3}$, David Jin$^{4}$, \\ \large{\textbf{Zhiyuan Liu$^{1}${$^*$}\thanks{{$^*$}~~Corresponding authors.}, and Zhenghao Liu$^{5}${$^*$}}} }

\address{$^{1}$NLP Group, DCST, IAI, BNRIST, Tsinghua University, Beijing, China \\ 
$^{2}$School of Comp. Sci. \& Tech., Huazhong University of Science and Technology, Wuhan, China \\
$^{3}$Language Technologies Institute, CMU, Pittsburgh, PA, USA \\
$^{4}$Laboratory for Information and Decision Systems, MIT, Cambridge, MA, USA \\
$^{5}$Department of Computer Science and Technology, Northeastern University, Shenyang, China \\
         yus21@mails.tsinghua.edu.cn, facicofan@gmail.com, cx@cs.cmu.edu, \\
         jindavid@mit.edu, liuzy@tsinghua.edu.cn, liuzhenghao@mail.neu.edu.cn}

\input{abstract}

\begin{document}
\maketitleabstract

\input{introduction}

\input{related}

\input{Tables/overall}

\input{method}

\input{experiments_setup}

\input{experiments}

\input{conclusion}


\section*{Acknowledgements}

This work is supported by the National Key R\&D Program of China (No. 2022ZD0116312) and the National Natural Science Foundation of China (No. 62236004, No. 62206042).


\nocite{*}
\section{Bibliographical References}\label{sec:reference}

\bibliographystyle{lrec-coling2024-natbib}
\bibliography{anthology,custom}

\clearpage
\appendix

\input{appendix}

\end{document}

%% file: abstract.tex
\abstract{
Common document ranking pipelines in search systems are cascade systems that involve multiple ranking layers to integrate different information step-by-step.
In this paper, we propose a novel re-ranker Fusion-in-T5 (FiT5), which integrates text matching information, ranking features, and global document information into one single unified model via templated-based input and global attention.
Experiments on passage ranking benchmarks MS MARCO and TREC DL show that FiT5, as one single model, significantly improves ranking performance over complex cascade pipelines. 
Analysis finds that through attention fusion, FiT5 jointly utilizes various forms of ranking information via gradually attending to related documents and ranking features, and improves the detection of subtle nuances.
Our code is open-sourced at \url{https://github.com/OpenMatch/FiT5}.
 \\ \newline \Keywords{document ranking, attention, fusion} }

%% file: introduction.tex
\section{Introduction}

Document ranking in information retrieval (IR) uses signals from many sources: text matching between queries and documents~\cite{nogueira2019passage}, numerical ranking features~\cite{zhang2021learning}, and pseudo relevance feedback (PRF) from other retrieved documents~\cite{li2023pseudo}.
These signals capture different aspects of relevance and are currently modeled by different techniques, such as neural ranking~\cite{DBLP:journals/corr/MitraC17}, learning to rank (LeToR)~\cite{liu2009learning}, and query expansion~\cite{DBLP:journals/csur/CarpinetoR12}.

Current search systems often incorporate these ranking techniques by a cascade pipeline with multiple layers of ranking models, each capturing a certain type of ranking signals~\cite{yates2021pretrained,zhang2021learning,dai2018convolutional}.
For example, the retrieved documents can be first re-ranked by a BERT ranker for text matching~\cite{nogueira2019passage}, then a LeToR model to combine numerical features~\cite{zhang2020selective,zhang2021learning,Zhang2022HLATREM,dai2018convolutional}, and finally re-ranked again by matching with the query enriched by top-ranked global documents~\cite{zheng2020bert,yu2021improving,li2023pseudo}.
Effective it is, the multi-layered cascade pipeline is complicated to tune, and the barrier between layers inevitably restricts the optimization of search relevance.

In this paper, we introduce Fusion-in-T5 (FiT5), a T5-based~\cite{2019Exploring} ranking model that re-ranks documents within a unified framework using attention fusion mechanism. 
FiT5 is designed to consolidate multiple signals, including text matching, ranking features, and global document information, into a single, simple model.
Specifically, we pack the input to FiT5 using a template that incorporates the document text with the ranking feature.
Furthermore, we introduce global attention layers on the representation tokens from the late layers of FiT5 encoders, enabling FiT5 to make comprehensive decisions by considering the collective information across top-ranked documents. 
With such a design, FiT5 can integrate all aforementioned types of signals and naturally learn a unified model through end-to-end training.

Experimental results on widely-used IR benchmarks MS MARCO~\cite{nguyen2016ms} and TREC DL 2019 \& 2020~\cite{craswell2020overview,2021Overview}, show that FiT5 exhibits substantial improvements over traditional re-ranking pipelines. 
On MS MARCO, FiT5 further outperforms Expando-Mono-Duo~\cite{pradeep2021expando}, a multi-stage re-ranking pipeline by 4.5\%.
Further analysis reveals that FiT5 effectively leverages ranking features through the attention fusion mechanism.
It can better differentiate between similar documents and ultimately produce a better ranking result.

%% file: related.tex
\section{Related Work}

\input{Figures/FiT5}

\paragraph{Cascade Document Ranking Pipeline}
A full cascade ranking pipeline may consist of one/multiple classical/neural ranker(s), LeToR model(s), and a stage of re-ranking with information from top-ranked documents.
Rankers include retrievers and re-rankers based on vocabulary matching (e.g. BM25) or neural networks.
Neural rankers are composed of deep neural networks~\cite{knrm} or pre-trained language models~\cite{nogueira2019passage,yates2021pretrained}, optimized with large amounts of data.
A Learning-to-Rank (LeToR) model, such as a linear combination model~\cite{metzler2007linear} or neural network~\cite{han2020learning,burges2005learning}, utilizes machine learning to generate a relevance score by considering ranking features extracted from the data or rankers.
Documents are finally re-ranked with collective information of all candidate documents, often accomplished by expanding the query with additional information via pseudo relevance feedback (PRF)~\cite{yu2021improving,li2023pseudo} or strengthening document-wise interaction through neural networks~\cite{pradeep2021expando,Zhang2022HLATREM}.
Though effective, these cascade methods require careful engineering and may be hard to optimize.

\paragraph{Attention Fusion over Multiple Text Sequences}
Fusion-in-Decoder (FiD)~\cite{fid} adds a T5 decoder model on top of multiple T5 document encoders to fuse multiple text evidences through the decoder-encoder attention and generate the answer for open-domain QA. 
Transformer-XH~\cite{zhao2020transformer-xh} builds eXtra Hop attention across the text evidences inside the BERT layers to model the structure of texts for multi-hop QA.
In this paper, we leverage the similar idea from Transformer-XH and propose attention fusion to incorporate variant ranking signals for the document ranking task.


%% file: Figures/FiT5.tex
\begin{figure}[!t]
	\centering
	\includegraphics[width=0.99\linewidth]{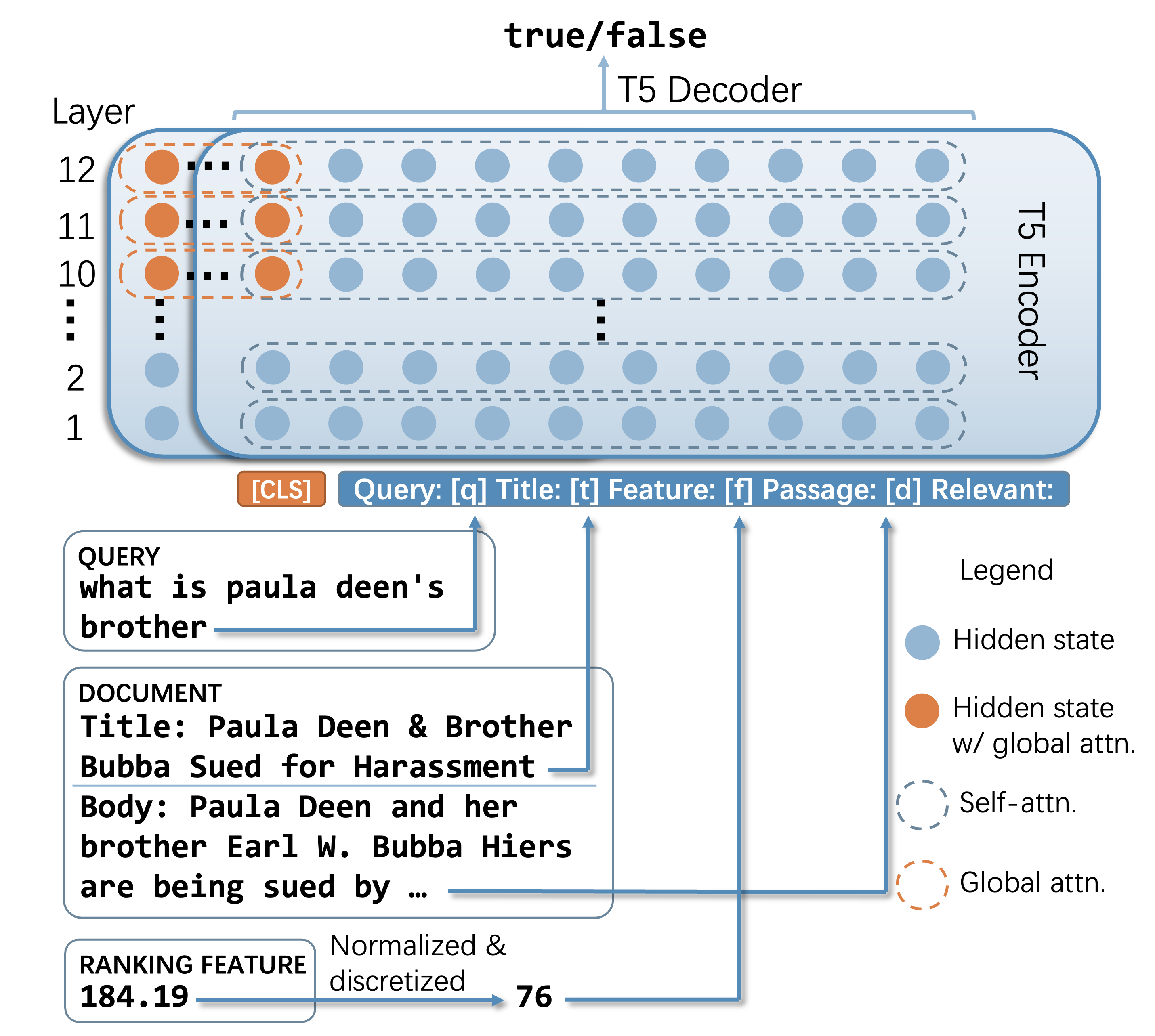}
	\caption{Architecture of Fusion-in-T5.
            The query, document, and ranking feature are filled in the template to form the input.
            In this paper, we use the retrieval score as the ranking feature.
 }
	\label{fig:framework}
\end{figure}

%% file: Tables/overall.tex
\begin{table*}
\centering
\small
\resizebox{0.99\linewidth}{!}{
\begin{tabular}{l|l|l|ll|ll|ll}
\hline
\multirow{2}{*}{\textbf{Model}}  & \multirow{2}{*}{\textbf{Re-ranker PLM(s)}}             & \multirow{2}{*}{\makecell{\textbf{\# Params of} \\ \textbf{Re-ranker(s)}}}  & \multicolumn{2}{c|}{\textbf{MS MARCO}} & \multicolumn{2}{c|}{\textbf{TREC DL'19}}   & \multicolumn{2}{c}{\textbf{TREC DL'20}}    \\ 
                        &                                               &                                                           & \textbf{MRR@10}        & \textbf{MAP@10}           & \textbf{NDCG@10}       & \textbf{MRR}               & \textbf{NDCG@10}       & \textbf{MRR}               \\
\hline
\multicolumn{9}{l}{\textit{First Stage Retrieval}}                                                                                                                  \\
\hline
BM25                    & n.a.                                            & n.a.                                                        & 18.7          &  19.5         &  50.6        & 70.4             & 48.0         & 65.9             \\
ANCE~\shortcite{xiong2020approximate}                    & n.a.                                            & n.a.                                                        & 33.0          &  --           &  64.8        & --             & 64.6         & --             \\
ANCE-PRF~\shortcite{yu2021improving}                    & n.a.                                            & n.a.                                                    & 34.4          &  --           &  68.1        & --             & 69.5         & --             \\
\underline{c}oCondenser~\shortcite{cocondenser} & n.a.                                & n.a.                                                        & 38.3          & 37.6          & 71.5        & 86.8             & 68.0          & 84.4             \\ 
\hline
\multicolumn{9}{l}{\textit{Two-stage Ranking (coCondenser $\rightarrow$ *) }}                                                                                       \\
\hline
\underline{B}ERT Re-ranker~\shortcite{nogueira2019passage} & BERT-base              & 110M                                                      & 39.2$\text{}^{c}$          &  38.6$\text{}^{c}$         & 70.1         & 83.8             &  69.2        & 82.3             \\
\underline{m}onoT5~\shortcite{monot5} & T5-base                                     & 220M                                                      & 40.6$\text{}^{cb}$          &  39.9$\text{}^{cb}$         & 72.6         & 84.8            & 67.7         & 85.1             \\
FiT5 (ours)             & T5-base                                       & 227M                                                      & \textbf{43.9}$\text{}^{cbm}$ & \textbf{43.3}$\text{}^{cbm}$ & \textbf{77.6}$\text{}^{cbm}$& \textbf{87.4}    & \textbf{75.2}$\text{}^{cbm}$& \textbf{85.5}    \\ 
\hline
\multicolumn{9}{l}{\textit{Multi($\ge$3)-stage Ranking (For Reference)}} \\
\hline
HLATR-base~\shortcite{Zhang2022HLATREM} & RoBERTa-base                  & 132M                                                      & 42.5          & --            & --            & --                & --            & --                \\
HLATR-large~\shortcite{Zhang2022HLATREM} & RoBERTa-large                & 342M                                                      & 43.7          & --            & --            & --                & --            & --                \\
Expando-Mono-Duo~\shortcite{pradeep2021expando} & 2$\times$T5-3B        & 2$\times$3B                                               & 42.0          & --            & --            & --                & 78.4         & 88.0             \\
\hline
\end{tabular}}
\caption{
Overall results on MS MARCO and TREC DL 19 \& 20.
Superscripts $c$, $b$, and $m$ indicate statistically significant improvements over \underline{c}oCondenser, \underline{B}ERT Re-ranker, and \underline{m}onoT5 (permutation test; $p < 0.05$).
Inapplicable and unavailable results are marked by  ``n.a.'' and ``--'', respectively.
}
\label{tab:overall}
\end{table*}

%% file: method.tex
\section{Methodology}

In this section, we first present the overview of FiT5 in §\ref{sec:overview}, then discuss the input and output format in §\ref{sec:input} and the attention fusion in §\ref{sec:global}.

\subsection{Task and Model Overview}\label{sec:overview}

Given a query $q$, a re-ranking model ranks a set of $n$ candidate documents $D=\{d_1,d_2,...,d_n\}$ from first-stage retrieval by assigning them with a set of scores $S=\{s_1,...,s_n\}$.
Traditional re-ranking model accomplishes the re-ranking task by making point-wise predictions, i.e. $S=\{s_1,...,s_n\}=\{f(q,d_1),...,f(q,d_n)\}$, with the query $q$, the document $d_i$ ($i=1,...,n$), and the model $f$.
FiT5 makes a more comprehensive prediction by deciding globally with more features, i.e. $S=\text{FiT5}(q,D,R)$, where $R=\{r_1,...,r_n\}$ is the set of ranking features for all documents.


FiT5 is based on the encoder-decoder model T5~\cite{2019Exploring}, as shown in Figure~\ref{fig:framework}. 
The encoder takes a triple of $(q, d_i, r_i)$ as the input. 
Attention fusion is introduced in the late layers of the encoder as global attention layers to incorporate signals from other documents in $D$.
The final ranking score $s_i$ is decoded from the decoder.


\subsection{Input and Output}\label{sec:input}

We pack $(q, d_i, r_i)$ using a template to form the input to FiT5.
The template consists of slots for input data and several prompt tokens, defined as
\begin{equation*}
\resizebox{.99\hsize}{!}{
\text{Query:}~\texttt{[q]}~\text{Title:}~\texttt{[t]}~\text{Feature:}~\texttt{[f]}~\text{Passage:}~\texttt{[d]}~\text{Relevant:},
}
\end{equation*}
where \texttt{[q]}, \texttt{[t]} and \texttt{[d]} are slots for text features, corresponding to the query $q$, the title and the body of the document $d_i$, respectively.
\texttt{[f]} is the slot for the feature $r_i$.
In this paper, we use the retrieval score as the ranking feature, after min-max normalization and discretization.

The model is fine-tuned to decode the token ``true'' or ``false'' according to the input.
During inference, the final relevance score is obtained from the normalized probability of the token ``true''.




\subsection{Attention Fusion via Global Attention}\label{sec:global}

The global document set $D$ and its feature set $R$ may contain valuable information for generating the score for every document $d_i$,
which cannot be captured via point-wise inference over the ``local'' information $(q, d_i, r_i)$~\cite{yu2021improving}.
To enhance the effectiveness of ranking, we propose attention fusion in FiT5 to enable the model to better comprehend and differentiate these documents with their features in the ranking process.


In FiT5, each $(q, d_i, r_i)$ pair first runs through $l-1$ transformer encoder layers independently, as in vanilla T5.
The attention fusion mechanism is enabled in every layer $j \geq l$. 
The representation of the first token $\text{[CLS]}$ (prepended to the input), denoted as $h_{i,\text{[CLS]}}^j \in \mathbb{R}^c$, is picked out from the normal self-attention:
\begin{equation}
\small
    h_{i,\text{[CLS]}}^j,\hat{\mathbf{H}}_i^j=\text{Transformer}(\mathbf{H}_i^{j-1}),
    \label{eq:cls}
\end{equation}
where $\hat{\mathbf{H}}_i^j$ denotes the remaining part of the hidden representation, $c$ is the hidden size and Transformer is the transformer layer. 
The representations of the first tokens from all $n$ encoders are then fed into a \textit{global} attention layer, allowing fusion over non-local information:
\begin{equation}
\small
\begin{aligned}
&\hat h_{1,\text{[CLS]}}^j,...,\hat h_{n,\text{[CLS]}}^j \\
=&\text{Global\_Attention}(h_{1,\text{[CLS]}}^j,...,h_{n,\text{[CLS]}}^j).
\label{eq:global}
\end{aligned}
\end{equation}
Finally, the globally-attended representation $\hat h_{i,\text{[CLS]}}^j$ is added back to the hidden representation: 
\begin{equation}
\small
    \mathbf{H}_i^j = [h_{i,\text{[CLS]}}^j+\hat h_{i,\text{[CLS]}}^j;\hat{\mathbf{H}}_i^j].
\end{equation}

In this way, the information from other documents in $D$ and features in $R$ is modeled in the representation of the [CLS] token and is then propagated to the following layer(s) in the encoder.

%% file: experiments_setup.tex
\section{Experimental Methodology}

In this section, we discuss our experimental setup.

\paragraph{Datasets and Metrics}
We train FiT5 on MS MARCO passage ranking dataset~\cite{nguyen2016ms} and evaluate it on its development set and TREC Deep Learning Tracks (TREC DL) 2019 \& 2020~\cite{craswell2020overview,2021Overview}.
MS MARCO labels are binary sparse labels (0/1) with often one positive document per query. 
TREC DL labels are dense judgments on a four-point scale from irrelevant (0) to perfectly relevant (3) and thus are more comprehensive~\cite{craswell2020overview,2021Overview}.
We report MRR@10, MAP@10 on MS MARCO, and NDCG@10, MRR on TREC DL.



\paragraph{Implementation}
We use T5-base model~\cite{2019Exploring} as the backbone of our model. 
Global attention modules are added starting from the third to last layer (i.e. $l = 10$) of the T5 encoder, implemented as standard multi-head attention with 12 attention heads. 
We re-rank the top 100 retrieved documents from coCondenser~\cite{cocondenser} and use coCondenser retrieval score as the ranking feature in the template defined in §~\ref{sec:input}. 
Specifically, we first normalize the coCondenser scores using min-max normalization and then discretize them into integers in $[0,100]$ to serve as the input.
We first train FiT5 without the features for 400k steps and then train it with the ranking feature for 1.5k steps to obtain the final model. 

\paragraph{Baselines}
We compare FiT5 with typical two-stage retrieve-and-rerank pipelines including BERT Re-ranker~\cite{nogueira2019passage} and monoT5~\cite{monot5}.
These re-rankers are trained to assign a score for each $(q,d_i)$ text pair individually.
The first-stage retrieval for such pipelines is kept the same as it for FiT5.
We also report the performance of multi($\ge$3)-stage ranking pipelines including HLATR~\cite{Zhang2022HLATREM}, a list-aware ranking pipeline and Expando-Mono-Duo~\cite{pradeep2021expando}, a sophisticated ranking system that employs pairwise comparison.
The performance of common first-stage retrieval models are also reported.

%% file: experiments.tex
\section{Evaluation Results}


This section presents the overall results of FiT5, and analyzes its effectiveness.

\subsection{Overall Performance}

The results of passage ranking on MS MARCO and TREC DL are presented in Table~\ref{tab:overall}.
By incorporating multiple types of ranking information, FiT5 greatly improves over the first-stage retrieval model coCondenser, and outperforms typical BERT Re-ranker and monoT5 that re-rank on top of the same retriever.
On MS MARCO, FiT5 further outperforms multi-stage ranking pipelines HLATR-large and Expando-Mono-Duo, which use significantly larger models (RoBERTa-large ~\cite{liu2019roberta} / $2\times$T5-3B) and more re-ranking stages.
Note that Expando-Mono-Duo is extremely computation-expensive as it requires pairwise inference of $n\times(n-1)$ times~\cite{pradeep2021expando}.

To study the efficiency of FiT5, we measure its inference time and GPU memory usage in comparison to monoT5 on the development set of MS MARCO.
As shown in Table~\ref{tab:efficiency}, compared to monoT5, FiT5 exhibits a mere 4.5\% increase in GPU memory usage and only a marginal increase in inference time.
This confirms the efficiency of FiT5's architecture, making it well-suited for practical applications.




\input{Tables/efficiency}

\input{Tables/ablation}

\input{Tables/layer_ablation}

\subsection{Ablation Study}

In this section, we first study the contribution of attention fusion in the effectiveness of FiT5.
The results are presented in Table~\ref{tab:ablation}.
When we exclude the feature score (FiT5 (w/o feature)) or global attention (monoT5 (w/ feature)), both scenarios result in a noticeable decline in performance.
Notably, monoT5 (w/ feature) does not exhibit a significant performance improvement over monoT5, indicating that the ranking feature can't be effectively captured straightforwardly in a vanilla transformer model.
Employing a linear combination of the re-ranker score and the feature still lags behind FiT5, revealing that the use of global attention is the key to effectively integrating the information from the retriever and other documents.

\input{Figures/distribute}



We then investigate the impact of the number of global attention layers on performance. 
We re-train FiT5 with top 1, 2, 3, 6, and 12 transformer layer(s) incorporated with global attention, respectively. 
Results in Table \ref{table;ablation} reveal that starting to integrate global attention from a late layer is an optimal choice. 
Starting the integration too early may make the optimization harder, whereas starting too late may provide insufficient paths for reasoning.

\subsection{Attention Pattern} \label{sec:attention_score}

In this experiment, we investigate the attention patterns within FiT5 and illustrate the distribution of global attention weights in Figure~\ref{fig:attn_dist}.
As shown in Figure \ref{fig:attention_b}, within the final layer, the attention values between the most relevant passages (labeled 3) are notably higher than those involving other passages. 
As shown in Figure~\ref{fig:attention_a}, with increasing layer depth, the general trend reveals a diminishing emphasis in attention values between the most relevant passages and other passages.
This attention pattern shows that as data traverses through multiple global attention layers, it fosters a stronger interaction among relevant documents, facilitating the distinction between positive and negative ones.

%% file: Tables/efficiency.tex
\begin{table}
    \centering
    \begin{tabular}{lll}
    \hline
                & \textbf{Time}  & \textbf{Memory}\\
    \hline
        monoT5  & 19m35s& 6088MiB\\
        FiT5    & 19m37s& 6362MiB\\
    \hline
    \end{tabular}
    \caption{Inference time and GPU memory usage of FiT5 and monoT5 on MS MARCO dev set, measured on a single NVIDIA A100 40G GPU with a batch size of 100 question-document pairs per step.
    }
    \label{tab:efficiency}
\end{table}

%% file: Tables/ablation.tex
\begin{table}[t]
\centering
\small
\resizebox{0.99\linewidth}{!}{
\begin{tabular}{llll}
\hline
\textbf{Model}                           &      \textbf{MARCO}        & \textbf{DL'19}        &  \textbf{DL'20}    \\ 
\hline
\underline{m}onoT5                          & 40.56             & 72.55             & 67.73\\
monoT5 (\underline{w}/ feature)               & 40.95$\text{}^{m}$             & 72.12             & 68.73\\
FiT5 (w/\underline{o} feature)                & 42.79$\text{}^{mw}$              & 74.94$\text{}^{w}$             & 70.02\\
FiT5 (\underline{l}inear combination)            & 43.59$\text{}^{mwo}$             & 75.41$\text{}^{mw}$             & 70.95$\text{}^{mw}$\\
FiT5                            & \textbf{43.93}$\text{}^{mwo}$    & \textbf{77.63}$\text{}^{mw}$    & \textbf{75.24}$\text{}^{mwol}$\\ 
\hline
\end{tabular}
}
\caption{Contribution of attention fusion. 
The evaluation metric is MRR@10 on MS MARCO and NDCG@10 on TREC DL.
(permutation test; $p < 0.05$)}
\label{tab:ablation}
\end{table}

%% file: Tables/layer_ablation.tex
\begin{table}[t]
\centering
\small
\resizebox{.99\linewidth}{!}{
\begin{tabular}{lll}
\hline
\textbf{Model}            & \textbf{FiT5 (w/o feature)}          & \textbf{FiT5}          \\ \hline
\underline{A}ll layers ($l=1$)       & 41.23         &  40.83             \\
Top-\underline{6} layers ($l=7$)    & 42.49$\text{}^{an}$         &  43.36$\text{}^{an}$            \\
Top-\underline{3} layers ($l=10$)    & 42.79$\text{}^{an}$         &  \textbf{43.93}$\text{}^{a621n}$              \\
Top-\underline{2} layers ($l=11$)    & \textbf{42.95}$\text{}^{an}$&  43.43$\text{}^{an}$              \\
Top-\underline{1} layer ($l=12$)    & 42.78$\text{}^{an}$         &  43.07$\text{}^{an}$              \\
\underline{N}o global attention         & 41.49         &  40.95              \\ \hline

\end{tabular}}
\caption{Performance on MS MARCO with global attention started to introduce at top-$k$ layers. 
The metric is MRR@10. 
(permutation test; $p<0.05$)}
\label{table;ablation}
\end{table}

%% file: Figures/distribute.tex
%
%


\begin{figure}[t]
    \centering
    \begin{subfigure}[t]{0.49\columnwidth}
        \centering
        \includegraphics[width=\linewidth]{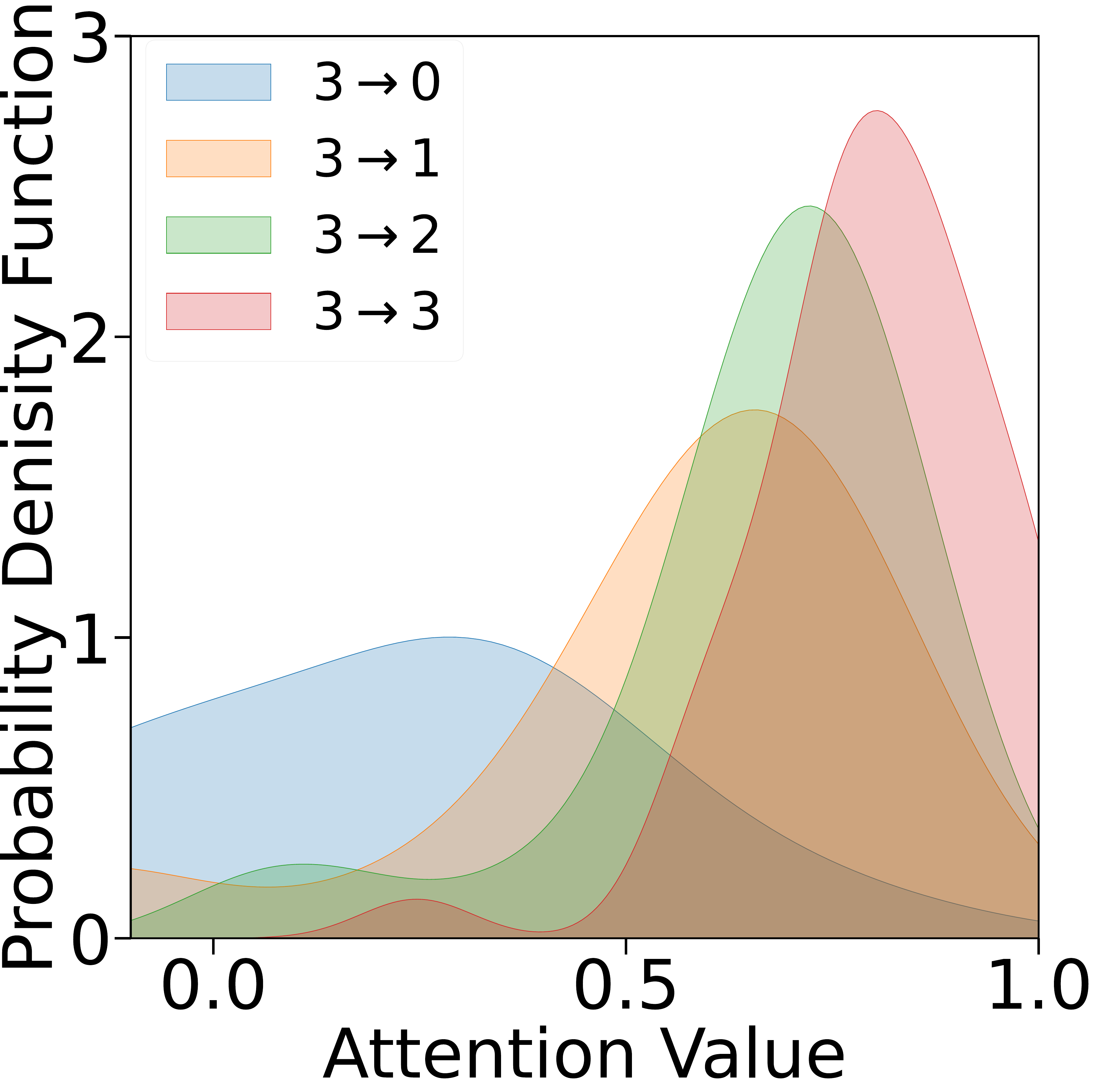}
        \caption{Attention between the most relevant passages (3) and passages in every relevance group (0--3). \label{fig:attention_b}}
    \end{subfigure}
    \begin{subfigure}[t]{0.49\columnwidth}
        \centering
        \includegraphics[width=\linewidth]{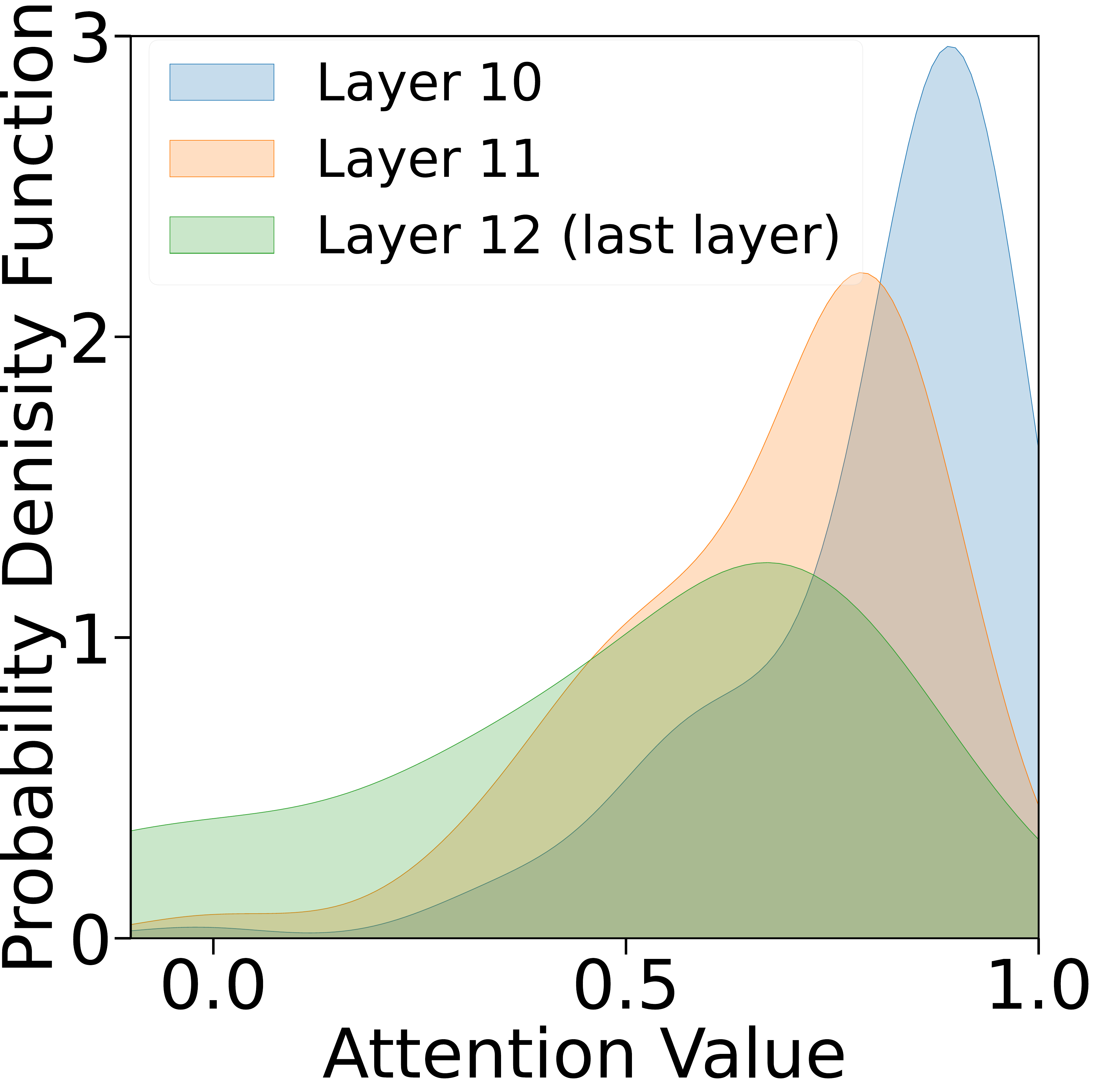}
        \caption{Attention between the most relevant passages and all others, grouped by layer. \label{fig:attention_a}}
    \end{subfigure}

    \caption{Attention weights distribution on TREC DL 20. 
    (a) depicts the distribution in the last layer (12); 0, 1, 2, and 3 are relevance levels from irrelevant to perfectly relevant.
    (b) presents the distribution in layer 10, 11, and 12. 
    }
    \label{fig:attn_dist}
\end{figure}

%% file: conclusion.tex
\section{Conclusion}
This paper introduces Fusion-in-T5 (FiT5), a unified ranking model that can capture variant information sources. 
It demonstrates superior or on-par results over cascade ranking systems with more stages.
Analysis reveals that the proposed attention fusion mechanism is effective in integrating signals including text matching, ranking features, and global information.

%% file: appendix.tex
\section{Ranking Feature Processing}\label{sec:feature}

We use the retrieval score from coCondenser~\cite{cocondenser} as the ranking feature. 
We first normalize the score to $[0,1]$ using min-max normalization. 
The scores are then discretized to an integer in $[0,100]$ by retaining two decimal places, and input to the model as normal strings.
The process can be formulated as:
\begin{equation}
    r = \text{floor}\left(\frac{\hat{r}-\hat{r}_\text{min}}{\hat{r}_\text{max}-\hat{r}_\text{min}} \times 100\right),
\end{equation}
where $\hat{r}$ denotes the raw score, $\hat{r}_\text{min}$ and $\hat{r}_\text{max}$ are the minimum and the maximum score, $\text{floor}\left(\cdot\right)$ is the flooring operation, and $r$ is the processed score used in the input.
In practice, we set $\hat{r}_\text{min}$ to 165 and $\hat{r}_\text{max}$ to 190.
Scores lower than $\hat{r}_\text{min}$ or greater than $\hat{r}_\text{max}$ are directly set to $\hat{r}_\text{min}$ or $\hat{r}_\text{max}$, respectively.

\section{Datasets}\label{sec:datasets}

\paragraph{MS MARCO Passage Ranking}~\cite{nguyen2016ms} is a ranking dataset with 8.8M short passages, constructed from Bing’s search query logs and web documents retrieved by Bing. 
The training and development split contains 530K and 6.9k queries, respectively.
We train FiT5 on the training split of MS MARCO. 
For every query, we take top-100 documents retrieved by coCondenser~\cite{cocondenser} for re-ranking. 
We use a held-out set of 3195 queries from the original training set for checkpoint selection and report the final results on the development set.
Note that we do not report the best results on the development set in our experiments.
We do not use the official test set as it requires submission to the leaderboard.

\paragraph{TREC Deep Learning Tracks}~\cite{craswell2020overview,2021Overview} are the test collections designed to study ad hoc ranking in a large data regime. 
TREC DL 2019 and 2020 contain 43 and 54 new test queries with human annotations, respectively.
The retrieval corpus is inherited from MS MARCO.
We follow the ``fullrank'' setup, where we directly perform retrieval from MS MARCO, not using the official first-stage retrieval results.

\section{Baselines}\label{sec:baselines}

We compare against the following baselines:

\paragraph{ANCE}~\cite{xiong2020approximate}
is a bi-encoder dense retrieval model based on RoBERTa-base.
It is trained iteratively on MS MARCO, first warmed-up using BM25 negatives and then trained using the hard negatives retrieved from the latest checkpoint.

\paragraph{ANCE-PRF}~\cite{yu2021improving}
is a bi-encoder dense retrieval model based on ANCE.
It further leverages pseudo relevance feedback (PRF) to refine the query representation.

\paragraph{coCondenser}~\cite{cocondenser}
is a bi-encoder dense retrieval model pre-trained using a Condenser task, which enhances the representation of the [CLS] by weakening the layer-wise connections of BERT, and a contrastive task.
It is then fine-tuned on MS MARCO.

\paragraph{BERT Re-ranker}~\cite{nogueira2019passage}
takes in a pair of $(q,d_i)$ and outputs a relevance score $s_i$ from the added linear head on the top.
We use BERT-base in our experiments.
We train BERT re-ranker on MS MARCO and re-rank the top 100 documents from coCondenser.
To have a fair comparison with FiT5, we also add the title to the input.
BERT re-ranker is trained using simple classification loss for 100k steps.

\paragraph{monoT5}~\cite{monot5}
is a point-wise re-ranking model based on the encoder-decoder model T5.
It is trained to output the token ``true'' or ``false'' to indicate relevance.
We use T5-base in our experiments.
We train monoT5 to re-rank the top 100 documents from coCondenser for 100k steps. 
Then, we continue training the model based on the previous checkpoints with the coCondenser retrieval score added as the ranking feature using the same template as FiT5 (Eq~\ref{eq:template}).
This becomes our monoT5 (w/ feature) run in our ablation study (Table~\ref{tab:ablation}).
To maintain consistency with FiT5, the title information is also added.

\paragraph{HLATR}~\cite{Zhang2022HLATREM}
adds an additional list-wise transformer-based re-ranking stage in the typical two-stage retrieve-and-rerank pipeline, combining first- and second-stage retrieval features.
The first stage is coCondenser retrieval and the second stage is a RoBERTa re-ranker.
The two variants HLATR-base and -large use RoBERT-base and -large re-ranker, respectively.
We directly report the results in their paper.
We refer readers to the original paper for more details.

\paragraph{Expando-Mono-Duo}~\cite{pradeep2021expando} is a series of multi-stage ranking pipelines.
On MS MARCO, we report the best-performing variant, which uses doc2query-T5~\cite{nogueira2019doc2query} to expand the document, BM25 as the first-stage retriever, monoT5-3B as the second-stage re-ranker, and duoT5-3B as the third-stage re-ranker.
On TREC DL 2020, we report the best-performing variant according to NDCG@10, which uses doc2query-T5 to expand the document, BM25 as the first stage retriever, RM3 pseudo relevance feedback, monoT5-3B as the second-stage re-ranker, and duoT5-3B as the third-stage re-ranker.
We refer readers to the original paper for more details.


\section{Implementation Details} \label{sec:training_detail}

To warm up for the final FiT5, we first train our model \textit{without} the feature, that is, to train with the following template:
\begin{equation}
\small
\text{Query:}~~\texttt{[q]}~~\text{Title:}~~\texttt{[t]}~~\text{Passage:}~~\texttt{[d]}~~\text{Relevant:},
    \label{eq:template_no_feature}
\end{equation}
for 400k steps, which results in the FiT5 (w/o feature) model in Table~\ref{tab:ablation}.
We then train it \textit{with} the ranking feature for 1.5k steps to obtain the final FiT5 model. 

In the training of FiT5 (w/o feature), the learning rate is $2\times 10^{-5}$, and the total batch size is 16. Each global attention module applies standard multi-head attention with 12 attention heads. We train the model for 400k steps on the MS MARCO and take the best-performing checkpoint on our held-out set. 
We then continue the training using the template with the feature for 1.5k steps to obtain the full FiT5 model. 
In the second training phase, the learning rate is $2\times 10^{-5}$, and the total batch size is 256 (with gradient accumulation).


In addition to incorporating feature information as text feature and fusing them with language model, we also employ a linear fusion method, shown in Table~\ref{tab:ablation} as FiT5 (linear combination). 
We use RankLib\footnote{https://sourceforge.net/p/lemur/code/HEAD/tree/RankLib/} to fuse the ranking score obtained from the first stage FiT5 (w/o feature) and the feature score from coCondenser. 
Specifically, we randomly sample 10k instances from the training data and train RankLib to obtain the linear fusion model, which is used as FiT5 (linear combination).

\input{Tables/layer_ablation_trecdl}
\section{Analysis Details}

In the experiment analyzing attention distribution in §\ref{sec:attention_score}, we compute attention values using the following method. We assume that the global attention similarity between the $i$-th and $k$-th samples in the $j$-th layer of transformers is denoted by $A_{i,k}^j$:
\begin{equation}
    A_{i,k}^j=\frac{\hat h_{i,\text{[CLS]}}^j \cdot \hat h_{k,\text{[CLS]}}^j}{||\hat h_{i,\text{[CLS]}}^j|| \cdot ||\hat h_{k,\text{[CLS]}}^j||}.
\end{equation}
Assuming the $i$-th sample is associated with a relevance label $l_i$ for query $q$, we compute the mean value of global attention similarity $A_{q}^j(R_1,R_2)$ in the $j$-th layer between samples with relevance scores $R_1$ and $R_2$:
\begin{equation}
    \hat{A}_{q}^j(R_1,R_2)=\frac{\sum_{i=1,l_i = R_1}^n \sum_{k=1,l_k = R_2}^n A_{i,k}^j}{\sum_{i=1,l_i = R_1}^n \sum_{k=1,l_k = R_2}^n 1}.
\end{equation}
To facilitate smoother visualization of the results for all queries, we perform min-max normalization on the those scores in the same layer $j$.
\begin{equation}
    \{A_{q}^j(R_1,R_2)\}=\text{Min-Max}(\{\hat{A}_{q}^j(R_1,R_2)\}).
\end{equation}
For $j$ equal to 12, with $R_1$ at 3 and $R_2$ ranging from 0 to 3, the results are shown in Figure ~\ref{fig:attention_b}.
For $j$ equal to 10, 11, and 12, with $R_1$ and $R_2$ ranging from 0 to 3, the results are presented in Figure ~\ref{fig:attention_a}. 

\section{Output Score Distribution}\label{sec:score_distribution}

\input{Figures/monot5_FiT5_distribute}

In Figure~\ref{fig:score_fit5_monot5}, we present the scores of documents with different labels.
FiT5 produces more distinguishable, non-binary distributions, indicating that it can better capture the nuances between similar documents.

\section{Case Study}\label{sec:case}

\input{Tables/case}

In this section, we show two winning examples of FiT5 in Table~\ref{tab:case}.

In the first case, when the user inquires about ``family feud'', monoT5 erroneously ranks a passage discussing a ``family affair'' as the top result. 
These occasional errors demonstrate the limitations of point-wise re-rankers at times. 
FiT5 successfully identifies and selects the passage related to the correct entity.

In the second case, the user seeks a comparison between Wi-Fi and Bluetooth. 
While monoT5 retrieves a passage that only addresses the speed aspect, partially addressing the query, FiT5 fetches a more comprehensive passage, offering a better fit for the user's requirements.

Overall, FiT5 outperforms monoT5 in capturing the relevance of documents to the user's query. 
It excels in determining which document is ``more relevant'' compared to others, enhancing the precision of document ranking.

%% file: Tables/layer_ablation_trecdl.tex
\begin{table}[t]
\centering
\small
\begin{tabular}{lll}
\hline
\textbf{Dataset}            & \textbf{DL'19}          & \textbf{DL'20}          \\ \hline
\underline{A}ll layers ($l=1$)       & 74.93$\text{}^{n}$         &  70.85$\text{}^{n}$            \\
Top-\underline{6} layers ($l=7$)    & \textbf{79.21}$\text{}^{an}$         &  73.93$\text{}^{an}$            \\
Top-\underline{3} layers ($l=10$)    & 77.63$\text{}^{an}$         &  \textbf{75.24}$\text{}^{a621n}$              \\
Top-\underline{2} layers ($l=11$)    & 77.74$\text{}^{an}$&  73.25$\text{}^{an}$              \\
Top-\underline{1} layer ($l=12$)    & 77.59$\text{}^{an}$         &  72.23$\text{}^{an}$              \\
\underline{N}o global attention         & 70.95         &  67.68              \\ \hline

\end{tabular}
\caption{FiT5's performance on TREC DL with global attention started to introduce at top-$k$ transformer layers. The metric is NDCG@10. (permutation test; $p<0.05$)}
\label{table;ablation_trecdl}
\end{table}

%% file: Figures/monot5_FiT5_distribute.tex
\begin{figure}[t]
    \centering
    \begin{subfigure}[t]{0.49\columnwidth}
        \centering
        \includegraphics[width=\linewidth]{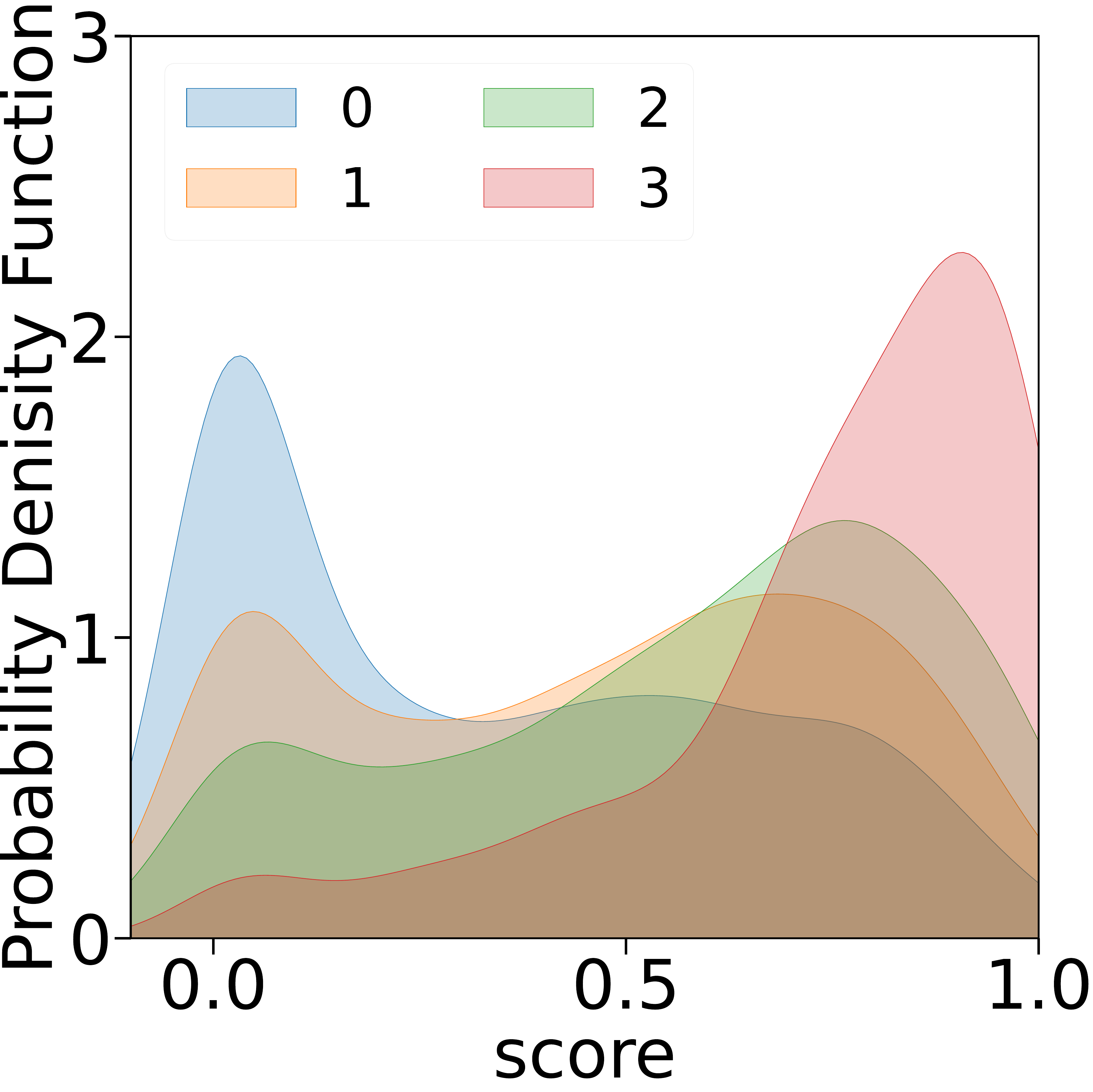}
        \caption{monoT5 (w/ feature). \label{fig:trecdl_score_a}}
    \end{subfigure}
    \begin{subfigure}[t]{0.49\columnwidth}
        \centering
        \includegraphics[width=\linewidth]{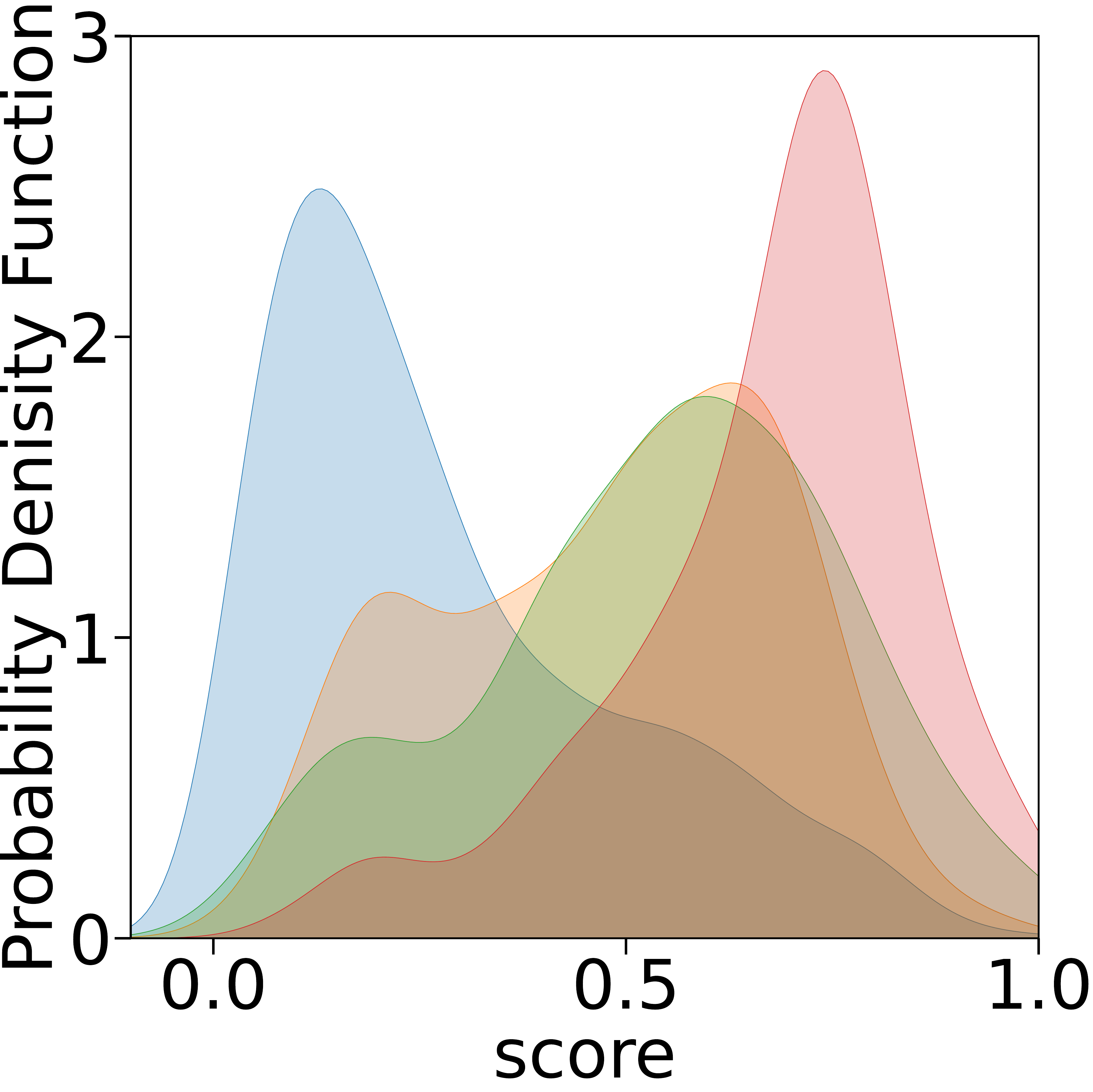}
        \caption{FiT5. \label{fig:trecdl_score_b}}
    \end{subfigure}
    \caption{Output score distributions on passages at different relevance levels from TREC DL 20.
    0, 1, 2, and 3 are relevance levels from irrelevant to perfectly relevant. 
    }
    \label{fig:score_fit5_monot5}
\end{figure}

%% file: Tables/case.tex
\begin{table*}[t]
\centering
\small
\resizebox{0.99\linewidth}{!}{
\begin{tabular}{l|c|c}
\hline
\hline & \textbf{FiT5} & \textbf{monoT5} \\
\hline \textbf{Query} & \multicolumn{2}{c}{when did family feud come out?}  \\
\hline \textbf{Title} & Richard Dawson & Family Affair \\
\hline \textbf{Snippet} & $\begin{array}{l}\text {... Dawson was hired by Goodson to host an  } \\
\text {upcoming project titled Family Feud, which} \\
\text {debuted on 12 July 1976 on ABC's daytime ...}\end{array}$ & $\begin{array}{l}\text { Family Affair is an American sitcom that aired } \\
\text {  on CBS from September 12, 1966, to March 4, } \\
\text { 1971.The series explored ... }\end{array}$ \\
\hline \textbf{Ranking Position} & 1 & 1 \\
\hline \textbf{TREC Label} & 3 (perfectly relevant) & 0 (irrelevant) \\
\hline \textbf{NDCG@10} &  75.62&34.27  \\
\hline
\hline \textbf{Query} & \multicolumn{2}{c}{what is wifi vs bluetooth}  \\
\hline \textbf{Title} & $\begin{array}{l}\text {   What is the difference between Bluetooth and} \\
\text {   Wi-Fi?    }\end{array}$  & $\begin{array}{l}\text {  Wi-Fi Direct vs. Bluetooth 4.0: A Battle for } \\
\text {   Supremacy    }\end{array}$ \\
\hline \textbf{Snippet} & $\begin{array}{l}\text {  ... The main difference is that Bluetooth is pri-   } \\
\text { marily used to connect devices without using    } \\
\text { cables, while Wi-Fi provides high-speed access } \\
\text { to the internet. ...} \end{array}$ & $\begin{array}{l}\text {  Bluetooth 4.0 vs. Wi-Fi Direct: Speed. Wi-Fi Di- } \\
\text {  rect promises device-to-device transfer speeds of } \\
\text { up to 250Mbps, while Bluetooth 4.0 promises sp- } \\
\text { eeds similar to Bluetooth 3.0 of up to ...}\end{array}$ \\
\hline \textbf{Ranking Position} & 1 & 1 \\
\hline \textbf{TREC Label} & 3 (perfectly relevant) & 2 (highly relevant) \\
\hline \textbf{NDCG@10} &  91.31&81.60  \\
\hline
\hline
\end{tabular}
}
\caption{Winning cases of FiT5 on TREC DL 2019.
We show the first passage that FiT5 and monoT5 disagree in the ranking results.}
\label{tab:case}
\end{table*}

%% file: emnlp2023.bbl
\begin{thebibliography}{41}
\expandafter\ifx\csname natexlab\endcsname\relax\def\natexlab#1{#1}\fi

\bibitem[{Burges et~al.(2005)Burges, Shaked, Renshaw, Lazier, Deeds, Hamilton,
  and Hullender}]{burges2005learning}
Chris Burges, Tal Shaked, Erin Renshaw, Ari Lazier, Matt Deeds, Nicole
  Hamilton, and Greg Hullender. 2005.
\newblock Learning to rank using gradient descent.
\newblock In \emph{Proceedings of ICML}, pages 89--96.

\bibitem[{Carpineto and Romano(2012)}]{DBLP:journals/csur/CarpinetoR12}
Claudio Carpineto and Giovanni Romano. 2012.
\newblock A survey of automatic query expansion in information retrieval.
\newblock \emph{{ACM} Comput. Surv.}, 44(1):1:1--1:50.

\bibitem[{Chen et~al.(2017)Chen, Fisch, Weston, and Bordes}]{chen2017reading}
Danqi Chen, Adam Fisch, Jason Weston, and Antoine Bordes. 2017.
\newblock Reading wikipedia to answer open-domain questions.
\newblock In \emph{Proceedings of ACL}, pages 1870--1879.

\bibitem[{Cormack et~al.(2009)Cormack, Clarke, and
  Buettcher}]{cormack2009reciprocal}
Gordon~V Cormack, Charles~LA Clarke, and Stefan Buettcher. 2009.
\newblock Reciprocal rank fusion outperforms condorcet and individual rank
  learning methods.
\newblock In \emph{Proceedings of SIGIR}, pages 758--759.

\bibitem[{Craswell et~al.(2021)Craswell, Mitra, Yilmaz, and
  Campos}]{2021Overview}
N.~Craswell, B.~Mitra, E.~Yilmaz, and D.~Campos. 2021.
\newblock Overview of the trec 2020 deep learning track.
\newblock In \emph{TREC}.

\bibitem[{Craswell et~al.(2020)Craswell, Mitra, Yilmaz, Campos, and
  Voorhees}]{craswell2020overview}
Nick Craswell, Bhaskar Mitra, Emine Yilmaz, Daniel Campos, and Ellen~M
  Voorhees. 2020.
\newblock Overview of the trec 2019 deep learning track.
\newblock In \emph{TREC}.

\bibitem[{Croft et~al.(2010)Croft, Metzler, and Strohman}]{croft2010search}
W~Bruce Croft, Donald Metzler, and Trevor Strohman. 2010.
\newblock \emph{Search Engines: Information Retrieval in Practice}, volume 520.
\newblock Addison-Wesley Reading.

\bibitem[{Dai et~al.(2018)Dai, Xiong, Callan, and Liu}]{dai2018convolutional}
Zhuyun Dai, Chenyan Xiong, Jamie Callan, and Zhiyuan Liu. 2018.
\newblock Convolutional neural networks for soft-matching n-grams in ad-hoc
  search.
\newblock In \emph{Proceedings of WSDM}, pages 126--134.

\bibitem[{Dalton et~al.(2019)Dalton, Xiong, and Callan}]{cast2019overview}
Jeffrey Dalton, Chenyan Xiong, and Jamie Callan. 2019.
\newblock Trec cast 2019: The conversational assistance track overview.
\newblock In \emph{Proceedings of TREC}.

\bibitem[{Gao and Callan(2022)}]{cocondenser}
Luyu Gao and Jamie Callan. 2022.
\newblock Unsupervised corpus aware language model pre-training for dense
  passage retrieval.
\newblock In \emph{Proceedings of ACL}, pages 2843--2853.

\bibitem[{Gao et~al.(2022)Gao, Ma, Lin, and Callan}]{Gao2022TevatronAE}
Luyu Gao, Xueguang Ma, Jimmy Lin, and Jamie Callan. 2022.
\newblock Tevatron: An efficient and flexible toolkit for dense retrieval.
\newblock \emph{arXiv preprint arXiv:2203.05765}.

\bibitem[{Han et~al.(2020)Han, Wang, Bendersky, and Najork}]{han2020learning}
Shuguang Han, Xuanhui Wang, Mike Bendersky, and Marc Najork. 2020.
\newblock Learning-to-rank with bert in tf-ranking.
\newblock \emph{arXiv preprint arXiv:2004.08476}.

\bibitem[{Izacard and Grave(2021)}]{fid}
Gautier Izacard and {\'E}douard Grave. 2021.
\newblock Leveraging passage retrieval with generative models for open domain
  question answering.
\newblock In \emph{Proceedings of EACL}, pages 874--880.

\bibitem[{Lee et~al.(2019)Lee, Chang, and Toutanova}]{lee2019latent}
Kenton Lee, Ming-Wei Chang, and Kristina Toutanova. 2019.
\newblock Latent retrieval for weakly supervised open domain question
  answering.
\newblock In \emph{Proceedings of ACL}, pages 6086--6096.

\bibitem[{Li et~al.(2023)Li, Mourad, Zhuang, Koopman, and
  Zuccon}]{li2023pseudo}
Hang Li, Ahmed Mourad, Shengyao Zhuang, Bevan Koopman, and Guido Zuccon. 2023.
\newblock Pseudo relevance feedback with deep language models and dense
  retrievers: Successes and pitfalls.
\newblock \emph{TOIS}, 41(3):1--40.

\bibitem[{Liu et~al.(2009)}]{liu2009learning}
Tie-Yan Liu et~al. 2009.
\newblock Learning to rank for information retrieval.
\newblock \emph{Foundations and Trends{\textregistered} in Information
  Retrieval}, 3(3):225--331.

\bibitem[{Liu et~al.(2019)Liu, Ott, Goyal, Du, Joshi, Chen, Levy, Lewis,
  Zettlemoyer, and Stoyanov}]{liu2019roberta}
Yinhan Liu, Myle Ott, Naman Goyal, Jingfei Du, Mandar Joshi, Danqi Chen, Omer
  Levy, Mike Lewis, Luke Zettlemoyer, and Veselin Stoyanov. 2019.
\newblock Roberta: A robustly optimized bert pretraining approach.
\newblock \emph{arXiv preprint arXiv:1907.11692}.

\bibitem[{Liu et~al.(2020)Liu, Xiong, Sun, and Liu}]{DBLP:conf/acl/LiuXSL20}
Zhenghao Liu, Chenyan Xiong, Maosong Sun, and Zhiyuan Liu. 2020.
\newblock Fine-grained fact verification with kernel graph attention network.
\newblock In \emph{Proceedings of ACL}, pages 7342--7351.

\bibitem[{Metzler and Bruce~Croft(2007)}]{metzler2007linear}
Donald Metzler and W~Bruce~Croft. 2007.
\newblock Linear feature-based models for information retrieval.
\newblock \emph{Information Retrieval}, 10:257--274.

\bibitem[{Mitra and Craswell(2017)}]{DBLP:journals/corr/MitraC17}
Bhaskar Mitra and Nick Craswell. 2017.
\newblock Neural models for information retrieval.
\newblock \emph{arXiv preprint arXiv:1705.01509}.

\bibitem[{Nguyen et~al.(2016)Nguyen, Rosenberg, Song, Gao, Tiwary, Majumder,
  and Deng}]{nguyen2016ms}
Tri Nguyen, Mir Rosenberg, Xia Song, Jianfeng Gao, Saurabh Tiwary, Rangan
  Majumder, and Li~Deng. 2016.
\newblock Ms marco: A human generated machine reading comprehension dataset.
\newblock \emph{choice}, 2640:660.

\bibitem[{Nogueira and Cho(2019)}]{nogueira2019passage}
Rodrigo Nogueira and Kyunghyun Cho. 2019.
\newblock Passage re-ranking with bert.
\newblock \emph{arXiv preprint arXiv:1901.04085}.

\bibitem[{Nogueira et~al.(2020)Nogueira, Jiang, Pradeep, and Lin}]{monot5}
Rodrigo Nogueira, Zhiying Jiang, Ronak Pradeep, and Jimmy Lin. 2020.
\newblock Document ranking with a pretrained sequence-to-sequence model.
\newblock In \emph{Findings of EMNLP}, pages 708--718.

\bibitem[{Nogueira and Lin()}]{nogueira2019doc2query}
Rodrigo Nogueira and Jimmy Lin.
\newblock From doc2query to doctttttquery.

\bibitem[{Nogueira et~al.(2019)Nogueira, Yang, Cho, and Lin}]{duobert}
Rodrigo Nogueira, Wei Yang, Kyunghyun Cho, and Jimmy Lin. 2019.
\newblock Multi-stage document ranking with bert.
\newblock \emph{arXiv preprint arXiv:1910.14424}.

\bibitem[{Pradeep et~al.(2021)Pradeep, Nogueira, and Lin}]{pradeep2021expando}
Ronak Pradeep, Rodrigo Nogueira, and Jimmy Lin. 2021.
\newblock The expando-mono-duo design pattern for text ranking with pretrained
  sequence-to-sequence models.
\newblock \emph{arXiv preprint arXiv:2101.05667}.

\bibitem[{Qiao et~al.(2019)Qiao, Xiong, Liu, and Liu}]{qiao2019understanding}
Yifan Qiao, Chenyan Xiong, Zhenghao Liu, and Zhiyuan Liu. 2019.
\newblock Understanding the behaviors of bert in ranking.
\newblock \emph{arXiv preprint arXiv:1904.07531}.

\bibitem[{Qu et~al.(2020)Qu, Yang, Chen, Qiu, Croft, and
  Iyyer}]{DBLP:conf/sigir/Qu0CQCI20}
Chen Qu, Liu Yang, Cen Chen, Minghui Qiu, W~Bruce Croft, and Mohit Iyyer. 2020.
\newblock Open-retrieval conversational question answering.
\newblock In \emph{Proceedings of SIGIR}, pages 539--548.

\bibitem[{Raffel et~al.(2020)Raffel, Shazeer, Roberts, Lee, Narang, Matena,
  Zhou, Li, and Liu}]{2019Exploring}
Colin Raffel, Noam Shazeer, Adam Roberts, Katherine Lee, Sharan Narang, Michael
  Matena, Yanqi Zhou, Wei Li, and Peter~J. Liu. 2020.
\newblock Exploring the limits of transfer learning with a unified text-to-text
  transformer.
\newblock \emph{J. Mach. Learn. Res.}, 21:140:1--140:67.

\bibitem[{Sun et~al.(2021)Sun, Qian, Liu, Xiong, Zhang, Bao, Liu, and
  Bennett}]{metaadaptrank}
Si~Sun, Yingzhuo Qian, Zhenghao Liu, Chenyan Xiong, Kaitao Zhang, Jie Bao,
  Zhiyuan Liu, and Paul Bennett. 2021.
\newblock Few-shot text ranking with meta adapted synthetic weak supervision.
\newblock In \emph{Proceedings of ACL}, pages 5030--5043.

\bibitem[{Vogt and Cottrell(1999)}]{vogt1999fusion}
Christopher~C Vogt and Garrison~W Cottrell. 1999.
\newblock Fusion via a linear combination of scores.
\newblock \emph{Information retrieval}, 1(3):151--173.

\bibitem[{Wu(2009)}]{wu2009applying}
Shengli Wu. 2009.
\newblock Applying statistical principles to data fusion in information
  retrieval.
\newblock \emph{Expert Systems with Applications}, 36(2):2997--3006.

\bibitem[{Xiong et~al.(2017)Xiong, Dai, Callan, Liu, and Power}]{knrm}
Chenyan Xiong, Zhuyun Dai, Jamie Callan, Zhiyuan Liu, and Russell Power. 2017.
\newblock End-to-end neural ad-hoc ranking with kernel pooling.
\newblock In \emph{Proceedings of SIGIR}, pages 55--64.

\bibitem[{Xiong et~al.(2021)Xiong, Xiong, Li, Tang, Liu, Bennett, Ahmed, and
  Overwikj}]{xiong2020approximate}
Lee Xiong, Chenyan Xiong, Ye~Li, Kwok-Fung Tang, Jialin Liu, Paul~N. Bennett,
  Junaid Ahmed, and Arnold Overwikj. 2021.
\newblock Approximate nearest neighbor negative contrastive learning for dense
  text retrieval.
\newblock In \emph{Proceedings of ICLR}.

\bibitem[{Yates et~al.(2021)Yates, Nogueira, and Lin}]{yates2021pretrained}
Andrew Yates, Rodrigo Nogueira, and Jimmy Lin. 2021.
\newblock Pretrained transformers for text ranking: Bert and beyond.
\newblock In \emph{Proceedings of WSDM}, pages 1154--1156.

\bibitem[{Yu et~al.(2021)Yu, Xiong, and Callan}]{yu2021improving}
HongChien Yu, Chenyan Xiong, and Jamie Callan. 2021.
\newblock Improving query representations for dense retrieval with pseudo
  relevance feedback.
\newblock \emph{arXiv preprint arXiv:2108.13454}.

\bibitem[{Zhang et~al.(2020)Zhang, Xiong, Liu, and Liu}]{zhang2020selective}
Kaitao Zhang, Chenyan Xiong, Zhenghao Liu, and Zhiyuan Liu. 2020.
\newblock Selective weak supervision for neural information retrieval.
\newblock In \emph{Proceedings of The Web Conference 2020}, pages 474--485.

\bibitem[{Zhang et~al.(2022)Zhang, Long, Xu, and Xie}]{Zhang2022HLATREM}
Yanzhao Zhang, Dingkun Long, Guangwei Xu, and Pengjun Xie. 2022.
\newblock Hlatr: enhance multi-stage text retrieval with hybrid list aware
  transformer reranking.
\newblock \emph{arXiv preprint arXiv:2205.10569}.

\bibitem[{Zhang et~al.(2021)Zhang, Hu, Liu, Fang, and Lin}]{zhang2021learning}
Yue Zhang, ChengCheng Hu, Yuqi Liu, Hui Fang, and Jimmy Lin. 2021.
\newblock Learning to rank in the age of muppets: Effectiveness--efficiency
  tradeoffs in multi-stage ranking.
\newblock In \emph{Proceedings of the Second Workshop on Simple and Efficient
  Natural Language Processing}, pages 64--73.

\bibitem[{Zhao et~al.(2020)Zhao, Xiong, Rosset, Song, Bennett, and
  Tiwary}]{zhao2020transformer-xh}
Chen Zhao, Chenyan Xiong, Corby Rosset, Xia Song, Paul Bennett, and Saurabh
  Tiwary. 2020.
\newblock Transformer-xh: Multi-evidence reasoning with extra hop attention.
\newblock In \emph{Proceedings of ICLR}.

\bibitem[{Zheng et~al.(2020)Zheng, Hui, He, Han, Sun, and
  Yates}]{zheng2020bert}
Zhi Zheng, Kai Hui, Ben He, Xianpei Han, Le~Sun, and Andrew Yates. 2020.
\newblock Bert-qe: Contextualized query expansion for document re-ranking.
\newblock In \emph{Findings of EMNLP}, pages 4718--4728.

\end{thebibliography}
